\documentclass[letterpaper, 10 pt, conference]{ieeeconf}  

\IEEEoverridecommandlockouts                              

\usepackage{graphicx} 
\usepackage{epsfig} 
\usepackage{mathptmx} 
\usepackage{times} 
\usepackage{amsmath} 
\usepackage{amssymb}  
\usepackage{booktabs} 
\usepackage{algorithm}
\usepackage{algpseudocode}
\usepackage{xcolor, colortbl}
\usepackage{balance}
\usepackage[
  colorlinks=true,
  linkcolor=blue,
  citecolor=blue,
  urlcolor=blue
]{hyperref}

\title{\LARGE \bf
Channel‑Selected Stratified Nested Cross‑Validation for \\ Clinically Relevant EEG‑Based Parkinson’s Disease Detection }
\author{
\authorblockA{
Nicholas R. Rasmussen,
Rodrigue Rizk,
Longwei Wang,
Arun Singh,
KC Santosh
} \\
\authorblockA{
USD AI Research, Department of Computer Science\\
University of South Dakota\\
Vermillion, SD 57069, USA\\
nicholas.rasmussen@coyotes.usd.edu,
\{rodrigue.rizk, longwei.wang, arun.singh, kc.santosh\}@usd.edu
}
}

\begin{document}

\maketitle
\thispagestyle{empty}
\pagestyle{empty}

\begin{abstract}

The early detection of Parkinson’s disease remains a critical challenge in clinical neuroscience, with electroencephalography (EEG) offering a non‑invasive and scalable pathway toward population‑level screening. While machine learning has shown promise in this domain, many reported results suffer from methodological flaws, most notably patient‑level data leakage, inflating performance estimates and limiting clinical translation. To address these modeling pitfalls, we propose a unified evaluation framework grounded in nested cross‑validation and incorporating three complementary safeguards: (i) patient‑level stratification to eliminate subject overlap and ensure unbiased generalization, (ii) multi‑layered windowing to harmonize heterogeneous EEG recordings while preserving temporal dynamics, and (iii) inner‑loop channel selection to enable principled feature reduction without information leakage. Applied across three independent datasets with a heterogeneous number of channels, a convolutional neural network trained under this framework achieved 80.6\% accuracy and demonstrated state‑of‑the‑art performance under held‑out population block testing, comparable to other methods in the literature. This performance underscores the necessity of nested cross‑validation as a safeguard against bias and as a principled means of selecting the most relevant information for patient‑level decisions, providing a reproducible foundation that can extend to other biomedical signal analysis domains.

\end{abstract}


\section{Introduction}

Parkinson’s disease (PD) is the second most common neurodegenerative disorder, affecting over 10 million people worldwide and imposing a substantial socioeconomic burden, yet clinical diagnosis typically relies on overt motor symptoms that emerge only after significant neuronal loss~\cite{li2025global, bloem2021parkinson, kalia2015parkinson}. This delay underscores the need to identify prodromal biomarkers—such as REM sleep behavior disorder and subtle motor changes—that may precede diagnosis by years and enable earlier intervention~\cite{bloem2021parkinson}. Recent advances in biomedical signal acquisition have made such early monitoring increasingly feasible, with EEG emerging as a particularly promising modality for non‑invasive assessment of prodromal and progressive PD symptoms due to its portability, low cost, and high temporal resolution~\cite{moreau2023overview, michel2019eeg}. Unlike MRI or PET, EEG enables scalable, repeated assessments, supporting longitudinal monitoring and biomarker discovery~\cite{michel2019eeg, cassani2018systematic}.

\begin{figure}[t!]
\centering
\includegraphics[width=.90\linewidth]{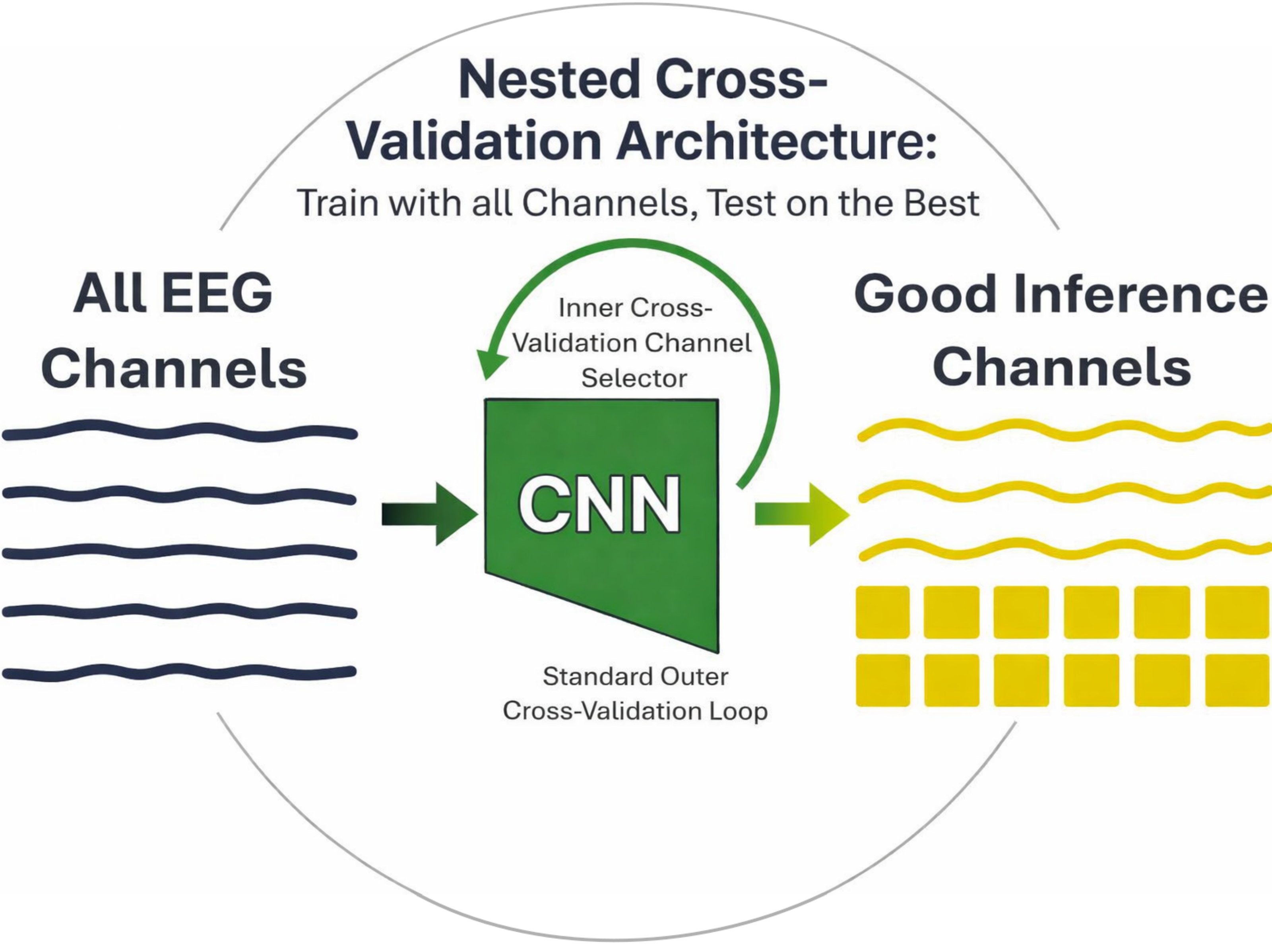}
\caption{The nested cross‑validation architecture is applied to high‑dimensional EEG by training models on all channels, while inner‑loop validation ranks channel relevance and defines the optimal test‑time architecture. The outer loop evaluates this selected configuration, enabling a dataset‑, channel‑, and model‑agnostic “many‑to‑best” reduction that supports efficient, generalizable, and interpretable inference.}
\label{fig:gF}
\end{figure}

In parallel, machine learning (ML) and deep learning (DL) methods show strong potential for extracting subtle biomarkers from high‑dimensional biomedical data~\cite{steyaert2023multimodal}. A wide range of neural architectures have been explored for biosignal analysis, including temporal recurrent models, self‑attention based transformer models, and graph neural networks that exploit EEG electrode layouts~\cite{roy2019deep, vafaei2025transformers}. Despite these advances, convolutional neural networks (CNNs) remain a principled starting point for EEG analysis. When applied to short‑time Fourier transform (STFT) spectrograms, CNNs exploit partial shift‑invariance aligned with the local stationarity of the STFT~\cite{hatami2018classification, chaman2021truly, 127284}. This synergy has consistently delivered state‑of‑the‑art performance across diverse signal domains, including PD detection, where CNN‑based or hybrid pipelines frequently report accuracies exceeding 95\%~\cite{LEE2021109282}. Moreover, CNN architectures are extensible to multimodal fusion, allowing EEG to be integrated with complementary biosignals through parallel branches~\cite{lee2025review}.

Many reported successes in EEG‑based PD detection are undermined by methodological issues, particularly data leakage and dataset heterogeneity. Leakage from overlapping temporal windows or subject overlap allows models to exploit patient‑specific idiosyncrasies rather than disease‑relevant features~\cite{torres2022methods}, and population‑level splits provide only partial protection against cohort‑specific overfitting~\cite{anjum2020linear}. Combined with heterogeneity in EEG channel counts, hardware, and preprocessing, these factors limit reproducibility and cross‑dataset transferability, motivating evaluation frameworks that explicitly prevent leakage and assess generalization across diverse populations.

To address these challenges, we propose a nested cross‑validation framework for robust EEG analysis (Figure~\ref{fig:gF}) that enforces patient‑level stratification in the outer loop to eliminate leakage, performs channel selection in the inner loop to reduce dimensionality without discarding clinically relevant signals, and integrates a multi‑layered windowing strategy to accommodate recordings with varying length, sampling rate, and hardware. By design, the framework is dataset‑, channel‑, and model‑agnostic, positioning it as a reproducible blueprint for EEG‑based biomarker discovery. Accordingly, our contributions are threefold: \emph{(i)} a model‑agnostic, stratified nested cross‑validation framework with integrated channel selection to prevent leakage, harmonize heterogeneous datasets, and reduce dimensionality; \emph{(ii)} empirical evidence that patient‑level leakage and narrow cohort blocking inflate reported EEG‑based PD detection performance and undermine clinical validity; and \emph{(iii)} demonstration that learned representations align with canonical EEG frequency bands, yielding interpretable models that achieve 80.6\% accuracy across datasets. 

Collectively, this framework supports clinical translation by enabling reproducible biomarker validation across diverse cohorts, facilitating integration with multimodal data (e.g., EMG and wearable sensors), and supporting scalable digital health pipelines for monitoring neurodegenerative disease progression in real‑world settings using ML.

\section{Related Work and Methodological Pitfalls}

A growing body of work applies ML techniques to early PD detection using EEG, EMG, and other biosignal modalities, with several studies reporting classification accuracies exceeding 85\%, suggesting strong translational potential~\cite{aljalal2022parkinson, khare2021detection}. However, many of these results rely on evaluation protocols that fail to enforce patient‑level stratification, allowing temporal windows from the same subject to appear in both training and test sets and introducing \textit{data leakage}. This enables models to exploit subject‑specific idiosyncrasies rather than learning generalizable disease markers, inflating reported performance estimates~\cite{torres2022methods, roberts2017cross, racine2000consistent, wu2024multi}.

Beyond data leakage, EEG‑based PD detection is limited by small sample sizes, accuracy‑only reporting, and preprocessing‑induced bias. Minor subject misallocation can substantially inflate performance~\cite{racine2000consistent}, while reliance on accuracy obscures clinically relevant trade‑offs~\cite{saito2015precision, steyerberg2010assessing}. Preprocessing choices further bias results when hyperparameters are tuned on overlapping data~\cite{roberts2017cross, valavi2018blockcv}. Although nested cross‑validation is standard for ensembling and optimization~\cite{roberts2017cross, schratz2019hyperparameter}, its use in PD detection would enable stability assessment across patient‑level splits and channel configurations, reducing overconfidence~\cite{wu2024multi}. Thusly, no existing studies evaluate non‑aligned cross‑dataset generalization, leaving cross‑protocol transferability unexplored.

\begin{figure*}[t!]
\centering
\includegraphics[width=.95\linewidth]{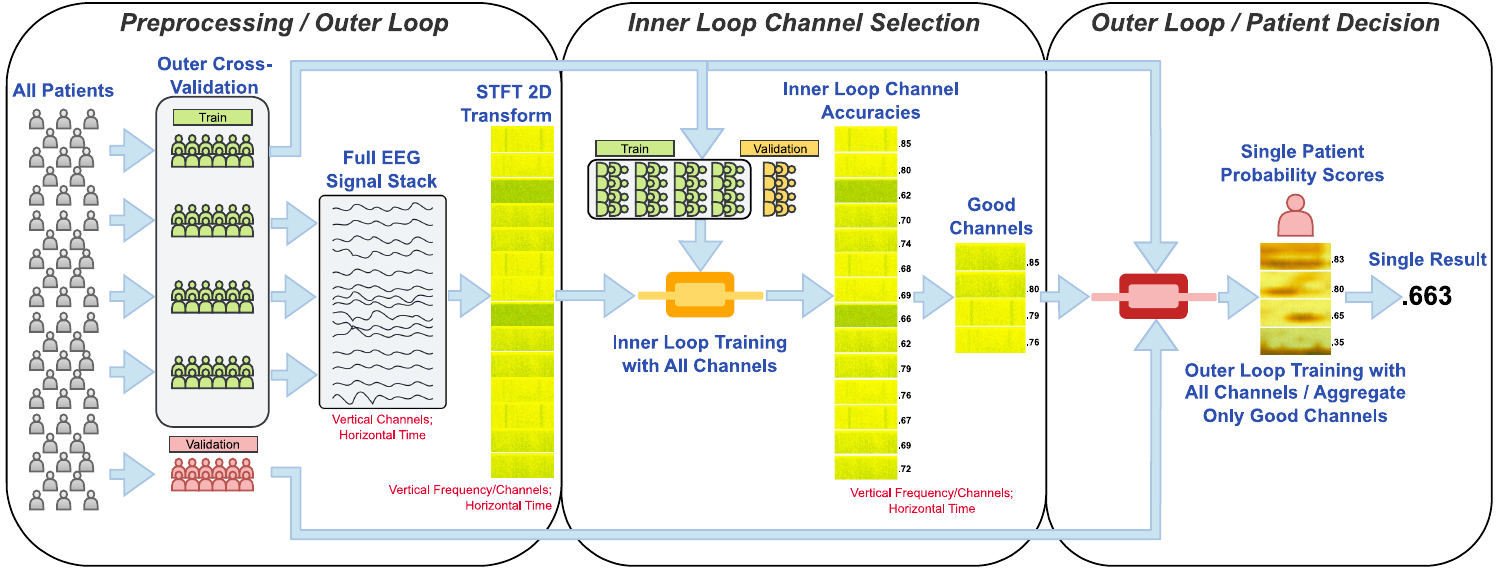}
\caption{Workflow for EEG channel selection: raw EEG is transformed into spectrograms, channels are ranked via inner‑loop cross‑validation, selected electrodes are evaluated in the outer loop, and window‑level predictions are aggregated into a single patient‑level probability.}
\label{fig:pipe}
\end{figure*}

Deep learning approaches for EEG‑based PD detection often inherit evaluation weaknesses, including non‑independent subject splits that inflate reported performance~\cite{oh2020deep, LEE2021109282}. Similar bias arises from non‑stratified cross‑validation and single‑population testing in later studies~\cite{anjum2020linear, shah2020dynamical, yuvaraj2018novel}. Although recent work explores cross‑dataset generalization via generative augmentation with reported accuracies near 85\%~\cite{zhang2025gepd}, such methods risk amplifying population‑specific artifacts when synthetic data reproduce dataset biases rather than disease‑relevant structure~\cite{williams2025eeg}.

Among recent studies, Wu et al.~\cite{wu2024multi} represent a methodological advance by explicitly addressing channel selection in EEG‑based PD detection. Their approach evaluates individual channels using a single train–validation split on the UNM dataset~\cite{anjum2020linear}, retaining electrodes that exceed a fixed performance threshold before retraining on the selected subset. While this provides insight into channel relevance and improves performance under population‑blocked testing, selection is performed once and remains fixed across runs, without accounting for variability from data partitioning or patient‑level heterogeneity. Moreover, the procedure is tightly coupled to a specific dataset and feature representation, limiting robustness across heterogeneous cohorts.

Our framework embeds channel selection within nested, patient‑stratified cross‑validation, replacing single‑pass, dataset‑specific procedures with stability‑adjusted estimates derived from repeated inner‑fold evaluations. All channels are retained during outer‑loop training to preserve shared spatial representations, while inference is restricted to the most reliable electrodes, yielding robust performance without leakage, singular validations, or dataset‑specific tuning and enabling more stringent, clinically relevant assessments of generalization under heterogeneous conditions.

\section{Proposed Nested Cross-Validation Architecture}

Our framework integrates preprocessing, windowing, and feature extraction with a \textit{nested cross-validation} design that enforces patient-level separation, supports principled channel selection, and, when desired, incorporates CNN-based band alignment for interpretability. The overall workflow is illustrated in Figure~\ref{fig:pipe} and narrated throughout this section.

\subsection{Preprocessing}
\label{M:Prep}
\paragraph{EEG Signal Preprocessing Steps} Let $\mathbf{s}_m(t)$ denote the raw EEG signal for patient $p_m$, sampled at frequency $f_{\text{orig}}$. Signals were resampled to $f_s = 64~\text{Hz}$ to preserve Beta ($13$–$30$~Hz) and Theta ($4$–$8$~Hz) activity, yielding $\tilde{\mathbf{s}}_m(t)$. Each trace was standardized,
\(
\mathbf{z}_m(t) = \frac{\tilde{\mathbf{s}}_m(t) - \mu_m}{\sigma_m},
\)
and normalized to unit amplitude,
\(
\hat{\mathbf{s}}_m(t) = \frac{\mathbf{z}_m(t)}{\max_t |\mathbf{z}_m(t)|}.
\)
Aligned signals $\hat{\mathbf{s}}_m(t)$ were stored with subject metadata to ensure traceability and leakage‑free partitioning across outer folds.

To ensure cross‑dataset comparability, channels were harmonized to a common spatial template. Let
\(
\mathcal{C} = \{c_1, c_2, \dots, c_C\}
\)
denote the full electrode set, with $|C|$ defined by the Iowa dataset. Channels were grouped into contiguous anatomical regions,
\(
\mathcal{C} = \mathcal{C}_{\text{frontal}} \cup \mathcal{C}_{\text{fronto‑central}} \cup \mathcal{C}_{\text{central}} \cup \cdots \cup \mathcal{C}_{\text{occipital}},
\)
and ordered within each region according to standard EEG conventions:
\begin{equation}
c \in \mathcal{C}_r =
\begin{cases}
\text{left hemisphere}, & \text{if index is odd}, \\
\text{midline}, & \text{if label ends in $z$},\\
\text{right hemisphere}, & \text{if index is even}. \\
\end{cases}
\label{eq:cases}
\end{equation}
Each channel was assigned an index
\(
\pi: \mathcal{C} \rightarrow \{0,1,\dots,C-1\},
\)
reflecting its left‑to‑right, anterior‑to‑posterior position. Channels absent from a recording were zero‑padded at index $\pi(c)$, ensuring all inputs conform to a consistent spatial structure across heterogeneous datasets.

\paragraph{Cross‑Validation Schema} We employ a patient‑stratified nested cross‑validation design in which the outer loop estimates generalization to unseen patients and the inner loop supports channel selection (Figure~\ref{fig:pipe}). Both loops use \textit{StratifiedGroupKFold} with patient identifiers as grouping variables, ensuring all samples $\mathcal{W}_m$ from a patient $p_m$ remain within a single fold. This structure prevents patient and temporal leakage, preserves class balance, and maintains unbiased outer‑loop evaluation by nesting channel selection within the inner loop.

\paragraph{Windowing and Feature Extraction}
To accommodate variable recording lengths and conditions, each patient’s EEG signal was segmented into channel‑wise spectrograms (Figure~\ref{fig:pipe}). With sampling rate $f_s = 64~\text{Hz}$, signals were windowed using
\(
L_w = 16{,}384~\text{samples} \; (\approx 256~\text{s}),
\)
with shorter recordings zero‑padded and longer recordings segmented using hop size
\(
H_w = \frac{L_w}{2} = 8{,}192~\text{samples}.
\)
Within each window, the short‑time Fourier transform (STFT) was computed. For a discrete‑time signal $\mathbf{w}[n] \in \mathbb{R}^{L_w}$,
\begin{equation}
\mathbf{STFT}(i,f) = \sum_{k=0}^{L-1} \mathbf{w}[bi + k] \cdot \mathbf{x}[k] \cdot e^{-j2\pi kf/L},
\label{eq:stft}
\end{equation}
where $L$ is the analysis window length, $b$ the hop size, $\mathbf{x}[k]$ the window function, $f$ the frequency bin index, and $i$ the time frame index. This yields a complex time–frequency representation indexed by $(i,f)$, with the number of time frames and frequency resolution given by
\(
T = \left\lfloor \frac{L_w}{b} \right\rfloor, \quad \Delta f = \frac{f_s}{L}.
\)
Using a discrete FFT implementation with $n_{\text{fft}} = 256$ and hop size $h = 64$, each channel produces magnitude spectrograms with
\(
F = \frac{n_{\text{fft}}}{2} = 128
\)
frequency bins and
\(
T = \frac{L_w}{h} = 256
\)
time frames, stored as
\(
(F \times T \times 1) = (128 \times 256 \times 1)
\)
tensors, converted to decibel scale and normalized.

\subsection{Inner Loop Channel Selection}

As shown in Figure~\ref{fig:pipe}, channel selection is embedded within the nested cross‑validation framework to address the high dimensionality of EEG recordings with $C \in \{32,64\}$ electrodes. For each outer‑fold training set, per‑channel accuracies $a_{c,f}$ are computed across inner‑folds $f \in \{1,\dots,F\}$ and averaged to obtain a stability‑adjusted score,
\begin{equation}
\bar{a}_c = \frac{1}{F}\sum_{f=1}^{F} a_{c,f}.
\label{eq:chanAcc}
\end{equation}
Channels are ranked by $\bar{a}_c$, and the top subset $\mathcal{C}^\ast$ is selected in the inner loop, ensuring dimensionality reduction is guided by cross‑validated evidence while preserving strict separation from outer‑fold evaluation preventing leakage.

\subsection{Patient Decision}
\paragraph{Adaptive Resolution Pooling Network (ARP‑N)} As illustrated in the `Single Patient Probability Scores' stage of Figure~\ref{fig:pipe}, we optionally employ the Adaptive Resolution Pooling Network (ARP‑N) \cite{rasmussen2025ecologically}. ARP‑N transforms variable‑sized spectrograms into consistent square feature maps through convolutional blocks, batch normalization, dropout, and specialized pooling, followed by a dense sigmoid output. For each channel $c$, the STFT spectrogram $\mathbf{X}_{c} \in \mathbb{R}^{F \times T}$ with $F=128$ frequency bins and $T=256$ time bins is adaptively pooled to $\mathbf{Z}_{c} \in \mathbb{R}^{8 \times 8}$. Along the frequency axis, \(
F = 128 \;\;\longrightarrow\;\; F' = 8,
\) thus spanning \begin{equation}
\Delta f' = \frac{32~\text{Hz}}{8} = 4~\text{Hz}.
\label{eq:bla}
\end{equation} This resolution aligns directly with canonical EEG bands—
\([0\!-\!4]\) Hz (Delta), \([4\!-\!8]\) Hz (Theta), \([8\!-\!12]\) Hz (Alpha), and \([12\!-\!30]\) Hz (Beta, subdivided)—embedding neuroscientific conventions into the architecture itself. In practice, ARP‑N highlights spectral-temporal regions most indicative of Parkinson’s Disease, providing implicit band‑level interpretability while reducing computational cost and improving generalization across heterogeneous datasets.

\paragraph{Patient‑Level Evaluation}
Evaluation is performed at the patient level to reflect clinical use. For each patient $p_m$, window‑level probabilities $\hat{y}_{mi}$ from spectrograms $\mathcal{W}_m$ are aggregated by averaging,
\begin{equation}
\hat{y}_m = \frac{1}{|\mathcal{W}_m|} \sum_{i=1}^{|\mathcal{W}_m|} \hat{y}_{mi},
\label{eq:patAg}
\end{equation}
yielding a single prediction per patient. This ensures performance reflects generalization across individuals rather than overlapping signal segments, aligning evaluation with the outer cross‑validation design.

\section{Experiments and Results}

\paragraph{Datasets} We combined three independent public EEG datasets to assess robustness and cross‑dataset generalization: (i) \textit{Iowa}~\cite{anjum2020linear}, consisting of 64‑channel resting‑state recordings acquired under eyes‑open conditions; (ii) \textit{University of New Mexico (UNM)}~\cite{anjum2020linear}, comprising multi‑session 64‑channel EEG from Parkinson’s disease patients and controls with both eyes‑open and eyes‑closed resting‑state segments recorded within the same session; and (iii) \textit{San Diego}~\cite{swann2015elevatedsynchrony}, a clinical 32‑channel EEG dataset reflecting hardware and montage variability. For datasets providing medication annotations (UNM and San Diego), only recordings acquired in the \textit{on‑medication} state were used.

\paragraph{Evaluation Metrics} With the dataset approximately balanced at the patient level, we report accuracy, area under the ROC curve (AUC), precision, and recall without adjustments. These metrics capture overall correctness, threshold discrimination, and the trade‑off between false positives and negatives, providing a robust evaluation of clinical relevance under balanced conditions for Parkinson's detection.

\paragraph{Implementation and Reproducibility} All experiments were implemented in Python using \texttt{tensorflow/keras}, with results logged to CSVs for reproducibility. Training used Adamax with exponential decay, early stopping, and checkpointing. Reproducibility is ensured by the nested cross‑validation channel‑selection procedure (Algorithm~\ref{alg:ncv}), which governs principled channel selection evaluation across folds reducing the feature set; see Algorithm~\ref{alg:ncv} and \textbf{https://github.com/USD-AI-ResearchLab/Nested-Channel-Selection} for the full specification and GitHub repository.

\paragraph{Our Results}
Table~\ref{tab:model_performance} summarizes performance across three evaluation paradigms. The first, \textit{No Stratification}, trains without patient‑level separation, permitting temporal and subject leakage; we also include results from Lee et al.~\cite{LEE2021109282} and Wu et al. \cite{wu2024multi}, who used a comparable setup. The second, \textit{Single‑Population Blocking}, reflects models optimized for a single site or cohort; our I‑4C model is shown alongside Anjum et al.~\cite{anjum2020linear} and Wu et al. \cite{wu2024multi} where they are evaluated with the same test set. On the other hand, Zhang et al.~\cite{zhang2025gepd} tests on the UNM dataset and is included to reinforce single hold-out population performance. The third, \textit{Stratified Cross‑Validation}, is our proposed framework, enforcing patient‑level separation across folds. Here, the \textit{All‑Channel} model uses all electrodes, while reduced‑channel variants (1‑, 2‑, 4‑, 8‑, 16‑Channel) demonstrate principled channel selection. These comparisons reveal inflated performance under non‑stratified or singular population‑blocked designs and more clinically relevant estimates with multi-site stratified cross‑validation schema, ensuring realistic estimates.

\begin{algorithm}[t!]
\scriptsize
\caption{ \scriptsize Nested Cross-Validation Channel Selection}
\begin{algorithmic}[1]
  \Require Full dataset $X_{data}$, labels $y_{data}$, patient groups $G_{data}$, channel indices $C_{data}$, parameters $\theta$, number of inner folds $K$, shared memory slots $L$
  \Ensure Best channel subset $\mathcal{C}^*$, performance across folds

  \State Initialize results list $\mathcal{R} \gets \emptyset$
  \State Initialize channel score dictionary $\mathcal{S}[ch] \gets \emptyset$

  \ForAll{outer fold $(X_{train}, X_{test})$}
    \State Store data into shared memory slots $L$
    \State Initialize inner cross-validation with $K$ splits

    \For{$k \gets 1$ to $K$}
      \State Split into inner-train and inner-validation sets:
      \Statex \hspace{\algorithmicindent} $(X^{(k)}_{tr}, y^{(k)}_{tr}, G^{(k)}_{tr}, C^{(k)}_{tr})$
      \Statex \hspace{\algorithmicindent} $(X^{(k)}_{val}, y^{(k)}_{val}, G^{(k)}_{val}, C^{(k)}_{val})$

      \State Store data into shared memory slots $L$
      \State $\theta \gets \texttt{runTrain}(L, \text{inner}=True)$
      \State $M \gets \texttt{runChan}(L, \theta)$

      \ForAll{channel $ch$ in $M$}
        \State Append accuracy $M[\text{Acc}][ch]$ to $\mathcal{S}[ch]$
      \EndFor
    \EndFor

    \State Compute channel accuracy: $\bar{a}[ch] \gets \text{mean}(\mathcal{S}[ch])$
    \State $\mathcal{C}^* \gets \text{Top-}m \text{ channels by } \bar{a}[ch]$
    \State Store $\mathcal{C}^*$ in $L$
    \State $\theta \gets \texttt{runTrain}(L, \text{inner}=False)$
    \State Record results restricted to $(\theta,\mathcal{C}^*)$ into $\mathcal{R}$
  \EndFor

  \State Aggregate results across folds into dataframe $D$
  \State Compute averages per configuration: $\bar{D} \gets \text{mean}(D)$
  \State \Return{$\mathcal{C}^*$, $\bar{D}$}
\end{algorithmic}
\label{alg:ncv}
\end{algorithm}

\subsection{Comparative Analysis}

The first paradigm, \textit{No Stratification}, achieved the highest apparent performance (Acc = .985, AUC = .986, Prec = .995, Rec = .978), consistent with results reported by Lee et al.~\cite{LEE2021109282} and Wu et al.~\cite{wu2024multi}. However, these scores reflect data leakage rather than genuine generalization, as patient overlap across folds inflates performance and near‑zero variance signals overfitting. We therefore include this paradigm only to illustrate how evaluation protocols that ignore temporal and patient‑level structure can yield misleading outcomes.

Under \textit{Single‑Population Blocking}, our 4‑Channel model achieved the strongest performance among evaluated methods. In the Iowa‑blocked evaluation, it attained Acc = .929, AUC = .929, Prec = .875, and Rec = 1.00, exceeding the mid‑.80s accuracies reported by Anjum et al.~\cite{anjum2020linear} and Wu et al.~\cite{wu2024multi} under comparable blocked designs. This suggests effective feature extraction and channel selection under single‑cohort tuning; however, restricting both training and testing to a single population remains vulnerable to site‑specific or acquisition‑dependent artifacts and does not ensure generalization beyond the evaluated cohort~\cite{zhang2025gepd}. Accordingly, strong performance under population‑blocked testing should be interpreted cautiously.

In contrast, our framework provides clinically realistic estimates of generalization on heterogeneous EEG datasets by enforcing patient‑level separation through \textit{Stratified Cross‑Validation}. Within this setting, the 4‑Channel configuration achieved the strongest overall trade‑off (Acc = .806 $\pm$ .092, Rec = .801 $\pm$ .134) while maintaining stable precision (.785 $\pm$ .159). The 2‑Channel model showed balanced but weaker performance, whereas the 8‑Channel variant favored precision at the expense of recall; the 16‑Channel model underperformed across all metrics, reflecting a trade‑off between spatial coverage and channel reliability under patient‑level evaluation. Importantly, channel selection is driven by stability‑adjusted inner‑fold performance: as the retained subset grows, marginal contributions diminish and variability across folds increases, leading to reduced generalization under patient‑level cross‑validation.

Taken together, these results show that inflated scores often arise from temporal or patient leakage, and that best‑case performance within a single dataset can obscure poor generalization under heterogeneous conditions. By contrast, stratified evaluation with principled channel selection yields more conservative but clinically meaningful estimates aligned with deployment‑ready generalization.

\subsection{Patient‑Level Aggregation Ablation}

Table~\ref{tab:aggregation_ablation} presents an ablation of patient‑level aggregation strategies within our nested cross‑validation framework. Across configurations, \emph{Mean} and \emph{Median} aggregation consistently produced the most balanced and stable outcomes, particularly for the 4‑Channel model, which achieved the strongest overall accuracy and recall. The 2‑Channel models showed similar aggregation behavior, indicating that aggregation strategies generalize across folds: \emph{Majority} performed competitively but favored precision, \emph{GMean} was modestly weaker, and \emph{Max} and \emph{Min} proved unstable, either amplifying outliers or collapsing recall. The \textit{Iowa Blocked Test} followed the same pattern, with \emph{Mean}, \emph{Median}, and \emph{Majority} yielding the most consistent results across folds.

Overall, these findings indicate that aggregation is a central design choice shaping clinical validity rather than a secondary detail. Robust rules such as \emph{Mean} and \emph{Median} stabilize fold‑to‑fold variance while preserving the sensitivity–specificity balance required for screening contexts. In contrast, unstable aggregation strategies exacerbate the same generalization failures observed under poor stratification or excessive channel counts. Accordingly, careful aggregation complements stratified cross‑validation and principled channel selection, completing our framework for reliable, clinically relevant, patient‑level evaluation.

\begin{table}[t!]
\caption{\scriptsize Performance under different stratification and channel configs.}
\centering
\scriptsize
\begin{tabular}{lcccccccc}
\toprule
Model      & Acc & $\sigma$ & AUC & $\sigma$ & Prec & $\sigma$ & Rec & $\sigma$ \\
\midrule
\multicolumn{9}{c}{\textbf{No Stratification}} \\
Wu \cite{wu2024multi}         & 1.00        & NA     & 1.00     &  NA   & 1.00  & NA     & 1.00       & NA      \\
Lee \cite{LEE2021109282} & .992      & NA     & .992     & NA    & .989       & NA      & .994       &  NA  \\
Ours          & .985      & 0     & .986     & .001    & .995       & .008      & .978       & .008      \\
\midrule
\multicolumn{9}{c}{\textbf{Single-Population Blocking}} \\
\textbf{I-4C (Ours)}        & .929       & NA     & 0.929     &  NA   & .875       & NA     & 1.00       & NA      \\
Anjum \cite{anjum2020linear}         & .857      & NA     & 0.852     &  NA   & .857       & NA     & .857       & NA      \\
Wu \cite{wu2024multi}         & .857        & NA     & .841     &  NA   & .916  & NA     & .786       & NA      \\
Zhang \cite{zhang2025gepd}         & .843        & NA     & NA     &  NA   & .840 & NA     & .840       & NA      \\
\midrule
\multicolumn{9}{c}{\textbf{Stratified Cross-Validation}} \\
All‑Channel  & .715 & .070 & .738 & .102 & .711 & .091 & .759 & .131 \\
1-Channel  & .752 & .124 & .797 & .130 & .777 & .158 & .714 & .136 \\
2-Channel  & .774 & .107 & .797 & .104 & .785 & .159 & .742 & .125 \\
\textbf{4-Channel}  & .806 & .092 & .799 & .067 & .785 & .159 & .801 & .134 \\
8-Channel & .767 & .117 & .752 & .154 & .819 & .108 & .706 & .157 \\
16-Channel & .748 & .092 & .753 & .143 & .793 & .112 & .703 & .154 \\
\bottomrule
\end{tabular}
\label{tab:model_performance}
\end{table}

\begin{table}[t!]
\caption{\scriptsize Ablation of aggregates for patient-level decisions.}
\centering
\scriptsize
\resizebox{\linewidth}{!}{%
\begin{tabular}{lcccccccc}
\toprule
Method      & Acc & $\sigma$ & AUC & $\sigma$ & Prec & $\sigma$ & Rec & $\sigma$ \\
\midrule
\multicolumn{9}{c}{\textbf{Single-Population Blocking}} \\
I-4C GMean  & .821 & NA & .913 & NA & .846 & NA & .786 & NA \\
\textbf{I-4C Majority}    & .929 & NA & .944 & NA & .875 & NA & 1.00 & NA \\
I-4C Max    & .607 & NA & .957 & NA & .560 & NA & 1.00 & NA \\
I-4C Mean   & .893 & NA & .944 & NA & .824 & NA & 1.00 & NA \\
\textbf{I-4C Median}    & .929 & NA & .944 & NA & .875 & NA & 1.00 & NA \\
I-4C Min    & .571 & NA & .832 & NA & 1.00 & NA & .143 & NA \\
\midrule
\multicolumn{9}{c}{\textbf{Stratified Cross-Validation}} \\
All GMean  & .685 & .103 & .734 & .095 & .608 & .153 & .777 & .143 \\
All Majority  & .715 & .070 & .738 & .102 & .711 & .091 & .759 & .131 \\
All Max  & .633 & .135 & .731 & .098 & .867 & .130 & .615 & .145 \\
All Mean  & .707 & .084 & .738 & .102 & .693 & .115 & .751 & .147 \\
All Median  & .715 & .070 & .734 & .101 & .711 & .091 & .759 & .131 \\
All Min  & .573 & .147 & .694 & .116 & .310 & .230 & .790 & .262 \\
2C GMean  & .785 & .092 & .806 & .086 & .785 & .159 & .766 & .130 \\
2C Majority  & .766 & .106 & .797 & .104 & .807 & .123 & .721 & .130 \\
2C Max  & .766 & .106 & .814 & .101 & .807 & .123 & .721 & .130 \\
2C Mean  & .774 & .107 & .797 & .104 & .785 & .159 & .742 & .125 \\
2C Median  & .774 & .107 & .801 & .104 & .785 & .159 & .742 & .125 \\
2C Min  & .749 & .091 & .806 & .087 & .721 & .187 & .750 & .126 \\
4C GMean  & .756 & .091 & .796 & .052 & .689 & .165 & .777 & .160 \\
\textbf{4C Majority}  & .801 & .084 & .799 & .067 & .829 & .099 & .762 & .141 \\
4C Max  & .765 & .100 & .816 & .072 & .829 & .099 & .711 & .129 \\
\textbf{4C Mean}  & .806 & .092 & .799 & .067 & .785 & .159 & .801 & .134 \\
\textbf{4C Median}  & .797 & .082 & .826 & .052 & .807 & .123 & .770 & .133 \\
4C Min  & .717 & .095 & .748 & .074 & .608 & .183 & .750 & .199 \\
\bottomrule
\end{tabular}
}
\label{tab:aggregation_ablation}
\end{table}

\begin{figure*}[t!]
\centering
\includegraphics[width=.98\linewidth]{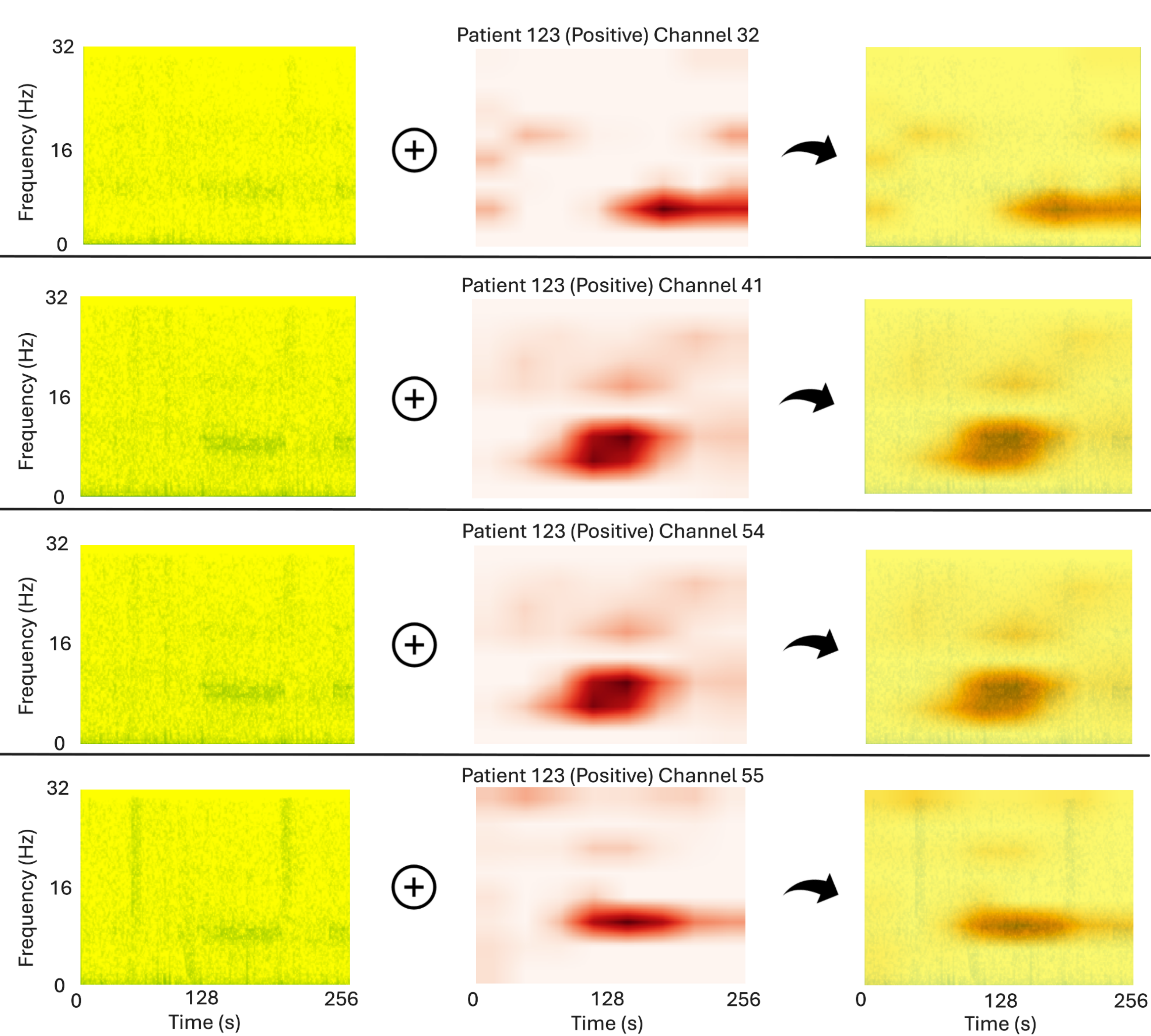}
\caption{Grad‑CAM overlays for Patient 123 highlight consistent band‑level focus. Each row shows one EEG channel (32, 41, 54, 55), progressing from spectrogram (yellow–green) to Grad‑CAM heatmap (red) to overlay. The model emphasizes recurrent theta activity (4–8 Hz), with secondary attention in alpha (8–12 Hz) and occasional beta (12–30 Hz). These theta–alpha patterns broaden and narrow, producing a pulsing appearance distinct from noise or artifacts. While edge effects (e.g., aliasing, zero‑padding) sometimes attract attention, the dominant focus remains on physiologically plausible oscillations.}
\label{fig:grad}
\end{figure*}

\subsection{Grad-CAM Interpretability Visualization}
Grad-CAM overlays in Figure~\ref{fig:grad} reveal that the network consistently emphasized the \textit{theta band (4--8 Hz)} across channels, with secondary contributions from \textit{alpha (8--12 Hz)} and \textit{beta (12--30 Hz)} ranges. In several cases, the model highlighted recurrent \textit{theta–alpha structures} with rhythmic fluctuations in spectral bandwidth, producing a pulsing appearance distinct from artifacts. While the CNN occasionally attended to \textit{processing artifacts} (e.g., zero-padding or aliasing), the dominant focus remained within physiologically meaningful bands; therefore aligning with canonical oscillatory rhythms suggesting learned representations are discriminative and \textit{neurophysiologically interpretable.} The frequency-specific spatially distributed activations across channels further indicate that the network captured \textit{cross-channel dependencies} rather than isolated channel effects.

This pattern aligns with prior EEG studies in PD, which report consistent alterations in theta and beta bands. For example, increased \textit{frontal theta activity} and impaired \textit{beta desynchronization} during lower-limb movement have been linked to motor deficits and freezing of gait (FOG) in PD patients \cite{singh2020frontal}. Complementary work using \textit{linear predictive coding} shows PD patients exhibit not only theta and beta abnormalities but also \textit{reduced alpha power}, distinguishing them from healthy controls \cite{anjum2020linear}. Thus, our Grad-CAM results suggest that ARP‑N is \textit{aligned with clinical research.} By discretizing the 0–32 Hz spectrum into 4 Hz bins, the network is structurally matched to canonical EEG bands, encouraging band‑level representations that Grad-CAM makes visible, thereby strengthening confidence in interpretability.

\section{Conclusion}

We introduced a nested cross-validation framework for early detection of PD from EEG, designed as a reproducible and generalizable response to common methodological pitfalls. By harmonizing three independent datasets (Iowa, UNM, San Diego) into a consistent montage, we established a foundation for cross‑site evaluation. Patient stratification in the outer loop prevented data leakage, while multi‑layered windowing preserved temporal structure and enabled uniform processing of heterogeneous recordings. The inner loop enabled principled channel selection, reducing dimensionality and improving generalization by emphasizing the most informative electrodes. Patient‑level aggregation, particularly mean, median, and majority, further reinforced clinical reliability. Collectively, these components show nested cross-validation as a principled model‑agnostic solution to leakage, dimensionality, and heterogeneity challenges that have historically inflated performance. Beyond PD, our architecture offers a reproducible framework for EEG and ML studies emphasizing scientific rigor and clinical applicability.

\section*{Acknowledgments}
\label{Ack}
This work was supported by the National Science Foundation under Grant No. \href{https://www.nsf.gov/awardsearch/showAward?AWD_ID=2346643}{\#2346643}, the U.S. Department of Defense under Award No. \href{https://dtic.dimensions.ai/details/grant/grant.14525543}{\#FA9550-23-1-0495}, and the U.S. Department of Education under Grant No. P116Z240151.
Any opinions, findings, conclusions or recommendations expressed in this material are those of the author(s) and do not necessarily reflect the views of the National Science Foundation, the U.S. Department of Defense, or the U.S. Department of Education. The author acknowledges Microsoft Copilot for assistance with grammar, structural edits, contextual extrapolation, and figure preparation, all with extensive human input \cite{CoPilot}. All core ideas, analyses, and interpretations are the author’s own based on previous literature and experimental analysis, as cited in the manuscript.

{\footnotesize
\bibliographystyle{IEEEtran}
\bibliography{root}
}

\addtolength{\textheight}{-12cm}   




\end{document}